# Einstein Gravitation Theory: Experimental Tests I

M. Cattani
Instituto de Fisica, Universidade de S. Paulo, C.P. 66318, CEP 05315-970
S. Paulo, S.P. Brazil . E−mail: mcattani@if.usp.br

Abstract.
 Using the Einstein gravitation theory (EGT) we calculate the Schwarzschild metric that is defined in the surrounding vacuum of a spherically symmetric mass distribution, not in rotation. The field equations of the EGT with this metric were applied to analyze the time dilation and the Doppler Effect of the light in order to test the validity of the EGT. This article was written to graduate and postgraduate students of Physics.

Resumo.
 Usando a teoria de gravitação de Einstein (TGE) nós calculamos a métrica de Schwarzschild que é definida no vácuo ao redor de uma distribuição esfericamente simétrica de massa, não em rotação. As equações de campo da TGE com essa métrica foram aplicadas para analisar a dilação temporal e o efeito Doppler luminoso afim de testar a validade da TGE. Esse artigo foi escrito para alunos de graduação e pós−graduação de Física.

**I. Introdução.**

Em artigos anteriores[1−4] mostramos como deduzir as equações de campo da teoria de gravitação de Einstein (TGE)[1] e, a partir delas, como calcular e detectar as ondas gravitacionais (OG).[2−4] Como as OG ainda não foram detectadas a TGE não pode ainda ser confirmada nesse contexto. No presente artigo usando a TGE calcularemos a métrica do espaço−tempo 4−dim que é gerada no vácuo ao redor de uma distribuição esfericamente simétrica de massa, não em rotação. Ela é conhecida como métrica de Schwarzschild.[5] Muitos fenômenos foram medidos e calculados[6−11] com muita precisão com a TGE usando essa métrica tais como: dilação temporal, deflexão e efeito Doppler da luz, precessão do periélio de planetas, precessão geodética de giroscópios e eco retardado de radar. Nesse artigo analisaremos somente a *dilação temporal* e o *efeito Doppler gravitacional da luz* visando testar a validade das previsões obtidas com as equações de campo da TGE. O nosso artigo visa o ensino de Física para alunos de Graduação e Pós−Graduação. Seguindo o procedimento adotado nos artigos anteriores[1−4] citaremos o menor número possível de referências, livros ou trabalhos originais.

# 1. Campo gravitacional esfericamente simétrico e estático.

Vejamos como calcular a métrica do espaço−tempo 4−dim de Riemann gerado por um campo gravitacional *estático* que envolve uma distribuição esfericamente simétrica de uma massa M em repouso e não em rotação. É óbvio que o campo deve ter simetria esférica. Por *estático* entendemos que além de independer do tempo, ou seja, além de ser *estacionário*, ele é temporalmente simétrico, ou seja, é invariante por uma inversão temporal. Vamos dar um exemplo da eletrodinâmica para entendermos isso.[5] Considere um campo magnético **B** produzido por um solenóide percorrido por uma corrente I constante. O campo é *estacionário*, isto é, independente do tempo. Entretanto, **B** não é *estático* pois, com uma inversão temporal t → −t a corrente I e, consequentemente, **B** invertem o sentido.

Para um campo com simetria esférica é natural[5] usarmos as coordenadas polares ct, r, θ e φ. Não há ambigüidade com θ e φ pois a medida dessas coordenadas depende da habilidade de dividirmos a circunferência concêntrica com a origem em partes iguais; isto pode ser feito sem o conhecimento da métrica. A coordenada radial r é ambígua pois não conhecemos ainda a sua relação precisa com a medida da distância. Assim, para começar, trataremos r simplesmente como um parâmetro que identifica diferentes superfícies esféricas concêntricas com a origem. A coordenada tempo t sofre do mesmo tipo de ambigüidade. Notemos que quando r → ∞ o espaço ficando plano os incrementos dr e dt se tornarão idênticos aos incrementos das distâncias e tempos reais[1−4], respectivamente.

O caráter estático do tempo impõe certas restrições sobre a forma do intervalo[1] $ds^2 = g_{\mu\nu}(x) dx^\mu dx^\nu$ do espaço−tempo. É fácil vermos que a solução estática proíbe termos do tipo dxdt, dydt e dzdt pois com uma inversão temporal esses termos mudariam de sinal. Desse modo os possíveis termos remanescentes seriam $dt^2$ e produtos formados por dx, dy e dz, dois a dois. Impondo ainda que haja simetria esférica o intervalo deverá ser uma função somente de combinações de x, y, z, dx, dy, dz que são invariantes por rotações espaciais. As únicas combinações possíveis seriam $(x^2+y^2+z^2)^{1/2} = (\mathbf{r}\cdot\mathbf{r})^{1/2}$, $dx^2+dy^2+dz^2 = d\mathbf{r}\cdot d\mathbf{r}$ e $xdx+ydy+zdz = \mathbf{r}\cdot d\mathbf{r}$. Considerando as coordenadas polares r, θ e φ as quantidades acima podem ficam escritas, respectivamente, como r, $dr^2 + r^2 d\theta^2 + r^2\sin^2\theta\, d\varphi^2$. Com essas restrições o intervalo $ds^2$, fazendo $x_o = x_4 = ct$, $x_1 = r$, $x_2 = \theta$ e $x_3 = \varphi$, seria dado por

$$ds^2 = A(r)c^2 dt^2 - B(r)(dr^2 + r^2 d\theta^2 + r^2 \sin^2\theta\, d\varphi^2) - C(r)dr^2 \qquad (1.1),$$

onde A(r), B(r) e C(r) são funções desconhecidas a serem determinadas resolvendo as equações de campo da TGE.[1,6−10] A (1.1) pode ser simplificada com uma transformação de coordenadas radial r´= r√B(r),

θ´ = θ , φ´=φ e t´= t, dando[5]

$$ds^2 = A'(r')c^2 dt'^2 - B'(r')dr'^2 - r'^2 d\theta'^2 - r'^2\sin^2\theta' d\varphi'^2$$

que pode ser reescrita de modo mais conveniente omitindo as linhas das coordenadas e colocando A´ e B´ como funções exponenciais:

$$ds^2 = e^{N(r)} c^2 dt^2 - e^{L(r)} dr^2 - r^2 d\theta^2 - r^2\sin^2\theta\, d\varphi^2 \qquad (1.2).$$

Desse modo, vemos que somente os elementos diagonais do tensor métrico $g_{\mu\nu}(x)$ são não nulos: $g_{oo} = g_{44} = c^2 e^{N(r)}$, $g_{11} = -e^{L(r)}$, $g_{22} = -r^2$ e $g_{33} = -r^2\sin^2\theta$. Com esses valores podemos calcular os símbolos de Christoffel definidos por [1,6–10]

$$\Gamma_{\mu\nu}{}^{\alpha} = \{{}_\mu{}^\alpha{}_\nu\} = (g^{\alpha\lambda}/2)(\partial_\nu g_{\lambda\mu} + \partial_\mu g_{\lambda\nu} - \partial_\lambda g_{\mu\nu}) \qquad (1.3),$$

lembrando que $g^{\mu\nu} = M_{\mu\nu}/|g|$ [9] onde g é o determinante de $g_{\mu\nu}$ e $M_{\mu\nu}$ é o determinante menor de $g_{\mu\nu}$ em g. Como os elementos de g são diagonais temos $|g| = |g_{11} g_{22} g_{33} g_{44}| = c^2 e^{N(r) + L(r)} r^4 \sin^2\theta$ obtemos $g^{oo} = c^{-2} e^{-N(r)}$, $g^{11} = e^{-L(r)}$, $g^{22} = r^{-2}$ e $g^{33} = r^{-2}\sin^{-2}\theta$. Como os $g_{\mu\nu}$ só dependem de $r = x_1$ em (1.3) só há derivadas do tipo $\partial_r g_{\mu\nu} = \partial_1 g_{\mu\nu}$. Indicando por $N' = \partial N/\partial r$ e $L' = \partial L/\partial r$ obtemos[5]

$$\Gamma_{o1}{}^o = \{{}_o{}^o{}_1\} = \Gamma_{1o}{}^o = \{{}_1{}^o{}_o\} = (N'/2), \qquad \Gamma_{oo}{}^1 = \{{}_o{}^1{}_o\} = (N'/2)e^{N-L},$$

$$\Gamma_{11}{}^1 = \{{}_1{}^1{}_1\} = L'/2, \qquad \Gamma_{23}{}^3 = \{{}_2{}^3{}_3\} = \Gamma_{32}{}^3 = \{{}_3{}^3{}_2\} = \cot\theta,$$

$$\Gamma_{12}{}^2 = \{{}_1{}^2{}_2\} = \Gamma_{21}{}^2 = \{{}_1{}^3{}_3\} = 1/r, \qquad \Gamma_{22}{}^1 = \{{}_2{}^1{}_2\} = -re^{-L}, \qquad (1.4)$$

$$\Gamma_{13}{}^3 = \{{}_1{}^3{}_3\} = \Gamma_{31}{}^3 = \{{}_3{}^3{}_1\} = 1/r, \qquad \Gamma_{33}{}^2 = \{{}_3{}^2{}_3\} = -\sin\theta \cos\theta \qquad \text{e}$$

$$\Gamma_{33}{}^1 = \{{}_3{}^1{}_3\} = -r \sin^2\theta\, e^{-L}$$

As equações de campo de Einstein são dadas por [1,6–10]

$$R_{\mu\nu} - (1/2) g_{\mu\nu} R = \kappa T_{\mu\nu}{}^{(m)} \qquad (1.5),$$

onde $\kappa = 8\pi G/c^4$, $R_{\mu\nu}$ é o *tensor de curvatura* de Ricci definido por

$$R_{\mu\nu} = R_{\nu\mu} = \partial_\mu \Gamma_{\nu\sigma}{}^\sigma - \partial_\sigma \Gamma_{\nu\mu}{}^\sigma + \Gamma_{\mu\sigma}{}^\tau \Gamma_{\nu\mu}{}^\sigma - \Gamma_{\nu\mu}{}^\tau \Gamma_{\tau\sigma}{}^\sigma \qquad (1.6)$$

e $R = g^{\mu\nu} R_{\mu\nu}$ é a *curvatura escalar* ou *invariante de curvatura*.[1,6–10]

Para determinarmos as funções L(r) e N(r) temos de resolver as equações (1.5) no vácuo onde $T_{\mu\nu}{}^{(m)} = 0$,

$$R_{\mu\nu} - (1/2)g_{\mu\nu} R = 0 \quad \text{ou} \quad R_\mu{}^\nu - (1/2)\delta_\mu{}^\nu R = 0 \quad (1.7),$$

onde $R_\mu{}^\nu = g^{\nu\alpha} R_{\alpha\mu}$ e $\delta_\mu{}^\nu = g^{\mu\nu} g_{\mu\nu}$ é o delta de Kronecker.

Substituindo os símbolos $\Gamma_{\mu\nu}{}^\alpha = \{{}_\mu{}^\alpha{}_\nu\}$ definidos por (1.4) em (1.6) obtemos os $R_{\mu\nu}$, calculamos R e através de uma elevação de índice $R_\mu{}^\nu = g^{\nu\alpha} R_{\alpha\mu}$ teremos, usando (1.7), as seguintes equações diferenciais

$$R_o{}^o - R/2 = -e^{-L}(L'/r - 1/r^2) - 1/r^2 = 0 \quad (1.8a)$$

$$R_1{}^1 - R/2 = -e^{-L}(N'/r + 1/r^2) - 1/r^2 = 0 \quad (1.8b)$$

$$R_2{}^2 - R/2 = R_3{}^3 - R/2 = e^{-L}[N''/2 - L'N'/4 + N'^2/4 + (N' - L')/2r] = 0 \quad (1.8c)$$

que podem ser facilmente integradas. Da (1.8a), por exemplo, decorre $e^{-L}(-rL' + 1) = 1$ que gera a solução única,

$$e^L = 1/(1 - a/r), \quad (1.9),$$

sendo a uma constante. Subtraindo (1.8a) de (1.8b) vemos que

$$L' = -N' \quad (1.10)$$

de onde deduzimos, com b = constante,

$$L(r) = -N(r) + b \quad (1.11).$$

Como $g_{11} = e^{L(r)} = 1/(1 - a/r)$ quando $r \to \infty$ verificamos que $g_{11} \to 1$ como é de se esperar para um espaço de Minkowski. Nesse mesmo limite como $g_{44} = g_{oo} = c^2 e^{N(r)} = c^2 e^{-L(r) - b}$ deve ser também igual a 1 verificamos que $c^2 e^{-b} = 1$. Desse modo teremos,

$$ds^2 = (1 - a/r)c^2 dt^2 - dr^2/(1 - a/r) - r^2 d\theta^2 - r^2 \sin^2\theta\, d\varphi^2 \quad (1.12).$$

Comparando $ds^2$ com o que se obtém para o caso de campos fracos[1,6–10] verificamos que $a = 2GM/c^2$. Resultando, finalmente,

$$ds^2 = (1 - 2GM/c^2 r) c^2 dt^2 - dr^2/(1 - 2GM/c^2 r) - r^2 d\theta^2 - r^2 \sin^2\theta\, d\varphi^2 \quad (1.13),$$

que é o resultado obtido por Schwarzschild em 1915.[5]

É sempre possível mudar a forma de $ds^2$ por uma transformação de coordenadas. Duas soluções diferentes de $ds^2$ são fisicamente idênticas se elas foram obtidas somente através de uma única transformação de coordenadas. Assim, ds² dado por (1.13) pode ser escrito de uma outra forma fazendo uma transformação de coordenadas $r \to 1 + GM/c^2r$,

$$ds^2 = \{(1-2GM/c^2r)/(1+2GM/c^2r)\}\, c^2dt^2 -$$

$$(1+2GM/c^2r)^4\,(dr^2 + r^2d\theta^2 + r^2\sin^2\theta\, d\varphi^2) \qquad (1.14).$$

De acordo com o *Teorema de Birkoff* [6,11] se para *qualquer tempo t* uma distribuição de massa for esfericamente simétrica a métrica do espaço−tempo gerada por ela no vácuo é idêntica a de Schwarzschild (1.13). Isso significa, por exemplo, que se durante um processo de contração ou colapso gravitacional de uma estrela a sua distribuição de massa se mantiver com simetria perfeitamente esférica a métrica de Schwarzschild é válida para uma região distante da estrela.

Para calcular ds² dado por (1.13) assumiu−se que a massa M *não estava em rotação*. A determinação da métrica levando em conta a rotação foi feita por Kerr.[12] Com a rotação não há simetria esférica; há uma simetria em torno do eixo de rotação z (Ohanian[6], pág. 324−326). A métrica é *estacionária* mas *não estática*: não depende do tempo porém não é invariante por uma inversão temporal devido a presença do termo $d\varphi\, dt$. A solução obtida por Kerr não é a única possível para um campo envolvendo uma massa em rotação. Diferentes soluções são obtidas dependendo das distribuições de massa, de seus momentos multipolos.[6]

Usando a TGE com métrica de Schwarzschild analisaremos a seguir a *dilação temporal* e o *Efeito Doppler da luz*.

## 2. Dilação temporal.

Conforme mostramos na Seção 1 a métrica do espaço−tempo gerada no vácuo em torno de uma distribuição esfericamente simétrica de uma massa M em repouso é definida, em coordenadas polares $x_o = x_4 = ct$, $x_1 = r$, $x_2 = \theta$ e $x_3 = \varphi$, através do invariante ds² dado por :

$$ds^2 = (1-2GM/c^2r)c^2dt^2 - dr^2/(1-2GM/c^2r) - r^2d\theta^2 - r^2\sin^2\theta\, d\varphi^2 \quad (2.1).$$

Consideremos dois relógios 1 e 2 em repouso nesse espaço com coordenadas de posição $r_1$ e $r_2$. Como para ambos temos $dr = d\theta = d\varphi = 0$ de (2.1) obtemos, sabendo que $ds^2 = c^2d\tau^2$ onde $\tau$ é o *tempo próprio*[1,6−11], ou seja, o tempo medido no referencial em que o relógio está em repouso,

$$d\tau_1 = \sqrt{g_{00}(1)}\, dt = (1-2GM/c^2r_1)^{1/2}\, dt$$

e
$$\qquad\qquad\qquad\qquad\qquad\qquad\qquad\qquad\qquad\qquad (2.2).$$

$$d\tau_2 = \sqrt{g_{00}(2)}\, dt = (1-2GM/c^2r_2)^{1/2}\, dt$$

É muito importante ressaltarmos que o tempo t que aparece em (2.1) e (2.2) é, no caso geral, uma *coordenada*, *não é* o tempo medido pelos relógios colocados no campo gravitacional. Somente no caso em que o relógio está muito longe dos efeitos gravitacionais, ou seja, quando r → ∞ o *tempo próprio* τ do relógio torna−se igual à *coordenada tempo* t. Segundo (2.2) um relógio situado a uma distância finita da origem mostra um intervalo de tempo dτ que é menor do que dt; isto é, a marcha do relógio é reduzida por um fator $\sqrt{g_{00}}$. Este efeito é chamado de *dilação temporal gravitacional*.

Estimando−se a perturbação gerada pela interação gravitacional no mecanismo responsável pelas transições atômicas que geram as freqüências usadas como padrões para medir o tempo em relógios atômicos (como o de Césio, por exemplo) verifica−se[12,13] que o efeito dessa perturbação é desprezível comparado com o devido ao fator $\sqrt{g_{00}}$. Nessas condições somos levados a concluir que a mudança na marcha do tempo é devida a um efeito puramente geométrico de *"encurvamento do espaço−tempo"*.

De modo análogo, no caso em que há somente uma alteração radial, ou seja, dr ≠ 0, dt = dθ = dφ = 0 de (1.1) obtemos, fazendo ds = dℓ

$$d\ell = \sqrt{g_{11}(r)}\, dr = dr/(1 - 2GM/c^2 r)^{1/2} \qquad (2.3),$$

onde r é a *coordenada* de um determinado ponto tomando o centro de M como origem. A distância entre dois pontos contíguos *não é* dr, *é* dℓ. Assim, verificamos que a separação dr entre as coordenadas é igual à distância dℓ entre os pontos somente quando r → ∞. A distância ℓ$_{12}$ entre dois pontos separados por coordenadas muito diferentes r$_1$ e r$_2$ é dada pela integral [1,11]

$$\ell_{12} = \int_{r_1}^{r_2} dr/(1 - 2GM/c^2 r)^{1/2} \qquad (2.4).$$

Desse modo, podemos concluir que a distância entre dois pontos ℓ$_{12}$ é diferente de r$_2$ − r$_1$ devido a um efeito puramente geométrico criado por um *"encurvamento do espaço−tempo"*.

Estimativas e medidas de dilação temporal foram efetuadas por Hafele[12] e Hafele e Keating[13]. As diferenças de tempo foram medidas com relógios atômicos, um em repouso na superfície da Terra do Observatório Naval (USA) e outro de Césio à bordo de um avião à jato realizando viagens de circunavegação em torno da Terra. O avião efetuou vôos, em sentidos leste−oeste e oeste−leste, a uma altura h da superfície da Terra com uma velocidade de cruzeiro V de um jato comercial. Para simplificar os cálculos vamos assumir que ambos os relógios estão ao longo do equador terrestre. A Terra gira em torno do eixo norte−sul geográfico que passa pelo seu centro de massa O tomado como a origem do sistema Σ (inercial) de coordenadas de posição (r,θ,φ). A velocidade angular de rotação da Terra em torno do eixo norte−sul será indicada por $\Omega = d\theta/dt$.

Usando (2.1) o relógio fixo na Terra, pondo $d\theta/dt = \Omega$, $r = R$ = raio da Terra, $d\varphi/dt = 0$, $dr/dt = 0$, $\chi = GM$ e M = massa da Terra, mede um intervalo tempo $d\tau_R$, dado por

$$d\tau_T^2 = \{(1- 2\chi/c^2 R) - R^2 \Omega^2/c^2\} dt^2 \qquad (2.5).$$

Para o relógio fixo no avião $r = R + h$, onde h é a altura do vôo, e que se move com velocidade V em relação ao solo a sua velocidade U em relação ao sistema inercial Σ será dada pela Relatividade Restrita (RR) por $U = [(R+h)\Omega +V]/[1+ (R+h)\Omega V/c^2]$. Supondo $c^2 \gg R\Omega V$ podemos fazer $U \approx (R+h)\Omega +V$. Desse modo, analogamente ao relógio fixo na Terra, o intervalo de tempo $d\tau_A$ medido no avião será dado por

$$d\tau_A^2 = \{ [1- 2\chi/c^2(R+h)] - [(R+h)\Omega +V]^2/c^2 \} dt^2 \qquad (2.6),$$

onde a velocidade tem sinais ± dependendo do sentido de movimento do avião em relação ao solo.

Antes de passarmos efetivamente ao cálculo de (2.5) e (2.6) é essencial definirmos as ordens de grandeza dos parâmetros R, h, χ/R, ΩR e V que são encontrados nos vôos de circunavegação: $M \approx 6 \cdot 10^{24}$ kg, $R \approx 6.4 \cdot 10^6$ m, $h \approx 10^4$ m, $V \approx 300$ m/s e $\Omega \sim 7.3 \cdot 10^{-5}$ rad/s. Assim, $\chi/R = GM/R \approx 6 \cdot 10^7$ Nm e $R\Omega \approx 460$ m/s. Desse modo como $|\chi/R| \ll c^2$, $V \ll c$, $R\Omega \ll c$ e $R \gg h$ obtemos, com um pouco de álgebra e considerando somente termos em primeira ordem, para um intervalo finito $\Delta t = \Delta t_o$,

$$\delta = \Delta\tau_A - \Delta\tau_T \approx \{ g^* h/c^2 - (2R\Omega + V)V/2c^2 \} \Delta t_o \qquad (2.7).$$

onde $g^* = GM/R^2 - R\Omega^2$ é a aceleração da gravidade no equador.

O primeiro termo de (2.7) é devido a um efeito "gravitacional" que só pode ser obtido com a TGE ao passo que o segundo termo é devido a um efeito "cinemático" que pode ser obtido usando somente a RR. O intervalo de tempo $\Delta t_o$ na circunavegação é assumido como sendo $\Delta t_o = 2\pi R/V$.

Medidas feitas em 1971[12,13] com relógios atômicos de referência do Observatório Naval (USA) e de Césio em vôos de circunavegação em jatos comerciais mostraram os seguintes resultados para δ, medido em ns,

$$\delta_{leste}(exp) = -59 \pm 10 \text{ ns} \quad \text{e} \quad \delta_{oeste}(exp) = 273 \pm 7 \text{ ns}$$

As correspondentes previsões teóricas são dadas por

$\delta_{leste}(teoria) = \delta_{grav} - \delta_{cinem} = [(144 \pm 14) - (184 \pm 18)]$ ns $= (- 40 \pm 23)$ ns

$$\delta_{oeste}(teoria) = \delta_{grav} - \delta_{cinem} = [(179 \pm 18) + (96 \pm 10)] \text{ ns} = (275 \pm 21) \text{ ns},$$

mostrando que há muito bom acordo entre teoria e experiência. Como a precisão dos relógios atômicos é da ordem[12,13] de $\delta \sim 10^{-13}$ s vemos que a TGE explica perfeitamente a *dilação temporal gravitacional*. Quando $\delta$ é negativo significa que o relógio que voa "anda" mais devagar do que o fixo na Terra e quando $\delta$ é positivo ele está "andando" mais depressa.

Dos resultados acima pode-se concluir que o $\delta$ *cinemático* é significativo, tão grande ou maior do que o gravitacional. De certo modo podemos dizer que esse fato é a resolução experimental do famoso *paradoxo dos relógios* (macroscópicos) *da RR*. Porém, com uma ressalva: os relógios estão fixos em sistemas acelerados, não inerciais.

Notemos que os efeitos de aceleração do avião durante a decolagem e a aterrissagem não foram mostrados explicitamente no cálculo de $\delta$ por serem desprezíveis comparados com os outros efeitos.[12,13]

Cutler[14] derivou uma relação mais geral para $\delta$ no caso de relógios circunavegando a Terra em uma latitude qualquer $\lambda$ ($\lambda = 0$ no equador). Ao invés de (2.7) obteve

$$\delta(\lambda) \approx \{ g^*h/c^2 - (2R\Omega \cos\lambda + V)V/2c^2 \} \Delta t_o.$$

Finalmente, como último comentário, como a Terra gira com velocidade angular $\Omega$ em torno do eixo z (Norte-Sul) a métrica do espaço-tempo deveria, rigorosamente, ser a de Kerr. Porém, os efeitos da rotação são desprezíveis[6,11] para $r \sim R$ se o fator $\alpha = GMa/c^2R^2 \ll 1$, onde $a = J_z/Mc$, $J_z = I\Omega$ é o momento angular da Terra ao longo do eixo z. Sendo $I \sim MR^2$ o momento de inércia da Terra em relação ao eixo z obtemos $\alpha \sim GM\Omega/c^2$. Como $M \sim 10^{24}$ kg e $\Omega \sim 7.3 \cdot 10^{-5}$/s temos $\alpha \sim 10^{-8} \ll 1$. Como nessas condições os efeitos da rotação podem ser desprezados a adoção da métrica de Schwarzschild é perfeitamente justificada para analisar fenômenos gravitacionais relativísticos gerados pela Terra.

## 3. Efeito Doppler da luz.

Conforme (2.2) dois relógios 1 e 2 com coordenadas $r_1$ e $r_2$ em repouso num campo gravitacional gerado por uma massa M "andam" com intervalos de tempos próprios $d\tau_1$ e $d\tau_2$ dados por

$$d\tau_1 = \sqrt{g_{00}(1)} \, dt = (1 - 2GM/c^2r_1)^{1/2} dt$$

e  (3.1).

$$d\tau_2 = \sqrt{g_{00}(2)} \, dt = (1 - 2GM/c^2r_2)^{1/2} dt$$

De (3.1) verifica−se que somente no caso em que o relógio está muito longe dos efeitos gravitacionais, ou seja, quando r → ∞ o *tempo próprio* τ do relógio torna−se igual à *coordenada tempo* t. Um relógio situado a uma distância finita da origem mostra um intervalo de tempo dτ que é menor do que dt; isto é, a marcha do relógio é reduzida por um fator $\sqrt{g_{00}}$. Este efeito é chamado de *dilação temporal gravitacional* e é devido a um efeito puramente geométrico de *"encurvamento do espaço−tempo"*. Conforme Seção 1 a dilação temporal foi confirmada através de experiências de circunavegação de relógios atômicos de Césio em jatos comerciais.[12,13]

Para entendermos o efeito Doppler gravitacional precisamos antes entender o princípio de funcionamento de um relógio atômico de Césio que foi usado por Hafele e Keating.[12,13] Através de microondas transições são induzidas entre dois níveis atômicos do Césio−133 que estão separados por uma diferença de energia hν onde ν = 9.192.631.770 Hz e h é a constante de Planck. Através de um mecanismo muito preciso[(*)] a freqüência ν é dividida por 9.192.631.770 para dar um pulso por *segundo*. Em outras palavras o *segundo*, pelo Sistema internacional de Unidades (SI) é definido como sendo o intervalo de tempo através do qual se tem 9.192.631.770 oscilações do átomo de Césio−133 exposto a uma excitação adequada. Ou seja, o intervalo de tempo é definido usando uma freqüência de transição entre dois níveis atômicos do Césio−133. Outros elementos químicos são utilizados para construir relógios atômicos mas, os mais usados são Hidrogênio, Rubídio e Césio.

Assim, conclui−se que a mudança da marcha do tempo é devida à alteração da freqüência de transição ν. Portanto, de (3.1) vemos que nos pontos $r_1$ e $r_2$ as freqüências de vibração devem ser $\nu_1 = 1/d\tau_1$ e $\nu_2 = 1/d\tau_2$, respectivamente, e ν = 1/dt. Conseqüentemente, de (2.1) decorre,

$$\nu_1 = \nu/(1-2GM/c^2 r_1)^{1/2} \quad \text{e} \quad \nu_2 = \nu/(1-2GM/c^2 r_2)^{1/2} \quad (3.2).$$

Usando (2.2) e assumindo que $r_{1,2} \gg 2GM/c^2$ teremos

$$(\nu_2 - \nu_1)/\nu = \Delta\nu/\nu \approx (GM/c^2)(1/r_2 - 1/r_1) \quad (3.3),$$

mostrando que para $r_2 \gg r_1$ a freqüência de ressonância em 2 é menor do que em 1, isto é, Δν < 0.

O cálculo do *efeito Doppler gravitacional* da luz pode realizado[6] usando diretamente a (3.3). Entretanto, calcularemos o efeito Doppler mais geral que é devido simultaneamente à gravitação e à cinemática. Para fazer isso precisamos considerar o movimento da luz que em TGE[1] se processa ao longo de uma geodésica nula. Assim, consideremos[9] um sistema emissor de luz em $r_1$ que está se movendo radialmente com velocidade $v_1$ e um detector em repouso em $r_2$. Suponhamos que 1 emita um fóton com freqüência $\nu_1$. Esse processo é definido pelo par de eventos $E_1$ e $E_1´$ cujas coordenadas são $(ct_1, r_1, \theta_1, \varphi_1)$ e $[c(t_1+ dt_1), r_1+ dr_1, \theta_1, \varphi_1]$ que geram o invariante $ds_1^2$ dado por,

$$ds_1^2 = (1- 2GM/c^2r_1)c^2dt_1^2 + dr_1^2/(1- 2GM/c^2r_1) =$$

$$= (1- 2GM/c^2r_1)(1- v_1^2/c^2)c^2dt_1^2 \qquad (3.4),$$

levando em conta que a velocidade de 1 é $v_1 = (dr_1/dt_1)/(1- 2GM/c^2r_1)$.[1,7,10]

A recepção do fóton pelo observador 2 é definida pelo par de eventos $E_2$ e $E_2´$ cujas coordenadas são $(ct_2, r_2, \theta_2, \varphi_2)$ e $[c(t_2+ dt_2), r_2, \theta_2, \varphi_2]$ que são separadas pelo intervalo

$$ds_2^2 = (1- 2GM/c^2r_2)c^2dt_2^2 \qquad (3.5).$$

Como os eventos $E_1$, $E_2$ e $E_1´$, $E_2´$ estão ao longo de uma geodésica do fóton[1,7,10] emitido que passa pelos pontos 1 e 2 temos, levando em conta que $ds = 0$:

$$dr/dt = c(1- 2GM/c^2r_1) \qquad (3.6).$$

Desse modo para o par $E_1$ e $E_2$ temos, integrando (3.6),

$$t_2 - t_1 = (1/c) \int_{r_1}^{r_2} dr/(1- 2GM/c^2r), \qquad (3.7)$$

e para o par $E_1´$, $E_2´$,

$$t_2+ dt_2 - (t_1+ dt_1) = (1/c) \int_{r_1+dr_1}^{r_2} dr/(1- 2GM/c^2r) \qquad (3.8).$$

Subtraindo (3.7) de (3.8) obtemos $dt_2 - dt_1 = -(1/c) dr_1/(1- 2GM/c^2r_1) = (v_1/c)dt_1$, ou seja,

$$dt_2 = (1 - v_1/c)dt_1 \qquad (3.9).$$

Dessa maneira de (3.4), (3.5) e (3.9) obtemos:

$$ds_2/ds_1 = [(1- 2GM/c^2r_2)/(1- 2GM/c^2r_1)]^{1/2} [(1- v_1/c)/(1+ v_1/c)]^{1/2} \qquad (3.10).$$

Com a (3.10) podemos calcular o *efeito Doppler* da luz levando em conta simultaneamente os efeitos gravitacional e cinemático. O *efeito Doppler gravitacional* é obtido colocando $v_1 = 0$ em (3.10). Neste caso re−obtemos (3.2) da qual deduzimos (3.3) quando $r_{1,2} >> 2GM/c^2$, fazendo $dt_1 = dt$ e $\nu = 1/dt$:

$$(v_2 - v_1)/v = \Delta v/v \approx (GM/c^2)(1/r_2 - 1/r_1) \quad (3.11).$$

A primeira medida feita no campo gravitacional da Terra testando a (3.11) foi realizada por Pound e Rebka em 1959.[15] Mediram a variação de freqüência de raios γ emitidos por núcleos de $Fe^{57}$ colocados no nível do solo e detectados por Efeito Mössbauer[16] a uma altura h = $r_2 - r_1$ = 22.6 m acima do solo. Eles mediram uma variação $(\Delta v/v)_{exp} \approx -2.46 \ 10^{-15}$, ou seja, um "redshift" do fóton ao subir uma altura h. Usando (3.11), fazendo $r_1$ = R e $r_2$ = R + h e levando em conta que R >> h a (3.11) fica dada por,

$$(\Delta v/v)_{teo} \approx -(g/c^2) h \quad (3.12),$$

onde g = $GM/R^2$ é aceleração gravitacional na Terra. Com a (3.12) verifica-se que $(\Delta v/v)_{exp}/(\Delta v/v)_{teo}$ = 1.05 ± 0.10 que mostra um ótimo acordo entre teoria e experiência. Numa versão mais moderna dessa experiência, feita por em 1964 por Pound e Snider,[17] constatou-se que $(\Delta v/v)_{exp}/(\Delta v/v)_{teo}$ = 1.00 ± 0.01.

*3.1 "Redshift" gravitacional em estrelas.*

Muitas medidas[6] foram feitas de "redshifts" gravitacionais de transições atômicas emitidas por átomos nas superfícies de estrelas. Assim, supondo que $M_*$ e $R_*$ sejam a massa e o raio da estrela, respectivamente, e que o observador esteja na Terra no ponto $r_2 >> R_*$ a (3.3) fica escrita como

$$\Delta v/v \approx -(GM_*/R_* c^2) \quad (3.13).$$

No caso do Sol, levando em conta que $G/c^2$ = 7.414 $10^{-28}$ m/kg, $M_*$ = 2.3 $10^{30}$ kg e $R_*$ = 1.394 $10^8$ m deveríamos ter, conforme (2.13), um "redshift" $(\Delta v/v)_{teo} \approx -2.12 \ 10^{-6}$. Esta previsão está em ótimo acordo com as medidas feitas por Brault[18] para o sódio e por Snider[19] para o potássio obtido $(\Delta v/v)_{exp}/(\Delta v/v)_{teo}$ = 1.0 ± 0.05 e $(\Delta v/v)_{exp}/(\Delta v/v)_{teo}$ = 1.01 ± 0.06, respectivamente, Observações de "redshifts" feitas para estrelas mais massivas[6] (Sirius A e B e Eridani B) e compactas como anãs brancas mostram que a (3.13) dá a ordem de grandeza correta para $(\Delta v/v)_{exp}$. Resultados semelhantes são obtidos para "redshifts" de raios-X emitidos por transições do Fe XXV e XXVI e do O VIII presentes na fotosfera de estrelas de nêutrons.[20,21]

Finalmente, ressaltamos que o redshift gravitacional (3.11) e (3.12) pode ser obtido de modo mais simples[6] usando: (a)conservação de energia e (b)princípio de equivalência substituindo o campo gravitacional por um campo de pseudo-força em um sistema de referência acelerado.

## 4. Conclusões

Do exposto nas Seções 2 e 3 concluímos, devido ao bom acordo entre os resultados experimentais e as previsões teóricas, que as equações de campo da TGE explicam de modo plenamente satisfatório a *dilação temporal* e o *Efeito Doppler da luz*.